# Biomechanics of Tori body motion during competition Sutemi (捨身技), Makikomi (巻込技), Tai Atari (体當り).

**Attilio Sacripanti**[1,2], **Envic Galea**[2,3], **Florin Daniel Lascau**[2,3]

1 University of Rome "Tor Vergata", Italy

2 IJF Academy Foundation, Malta

3 University of Hertfordshire, PhD Candidate, School of Life and Medical Science

**\* Corresponding author**: Attilio Sacripanti, Attilio.sacripanti@uniroma2.it.

**Authors Contribution Statement**
Professor Sacripanti developed the original concept and research design, which played a crucial role in its implementation, analysed the results, and edited the manuscript.
Dr. Galea analysed the results and revised the manuscript.
Dr. Lascau oversaw the project, focusing on elucidating the direction of falls (Ukemi), a crucial aspect in understanding the execution and safety of the throwing techniques being studied.
**Ethics Statement:** The Ethics Committee of the IJF Academy Foundation approved this study.
**Funding Statement:** The funders had no role in the study design, data collection and analysis, publication decisions, or manuscript preparation.
**Conflict of Interest Statement:** The authors declare no conflicts of interest regarding this manuscript

**Words Number**: 5527, with abstract 5778. (without References)
**Acknowledgements:** to Mr. Romeo Fabi for the valuable help in the layout

**Abstract:**
**Background/Objectives**: The biomechanics of Tori's body motion during Sutemi (捨身技), "sacrifice techniques", Makikomi (巻込) "wrapping techniques", and Tai Atari (体當り),"Body Strike techniques" in judo is crucial for understanding these techniques' effectiveness and safety. This study examined the biomechanics by analysing the entire body of the tori during execution.
**Methods**: We analysed Tori's body motion using original biomechanical models, considering the balance of sacrificing to throw Uke and body mass to strike the opponent. General equations for Sutemi, Makikomi, and Tai-atari were established, and their utilisation in high-level male competition was statistically analysed.
**Results**: The results show that effective Sutemi and Makikomi execution depends on the correct use of Tori's body mass. The biomechanical model revealed the relationships between angular falling speed, body angle, and athlete height. The model allowed us to evaluate the falling velocity, showing that the angular velocity and acceleration were proportional to the gravitational acceleration and body-Tatami angle and inversely proportional to the athlete height. The stopping and sliding forces are directly proportional to the sum of the Tori and Uke masses and the body-Tatami angle but are independent of the athlete's eight.
Tai Atari's force was quantified to evaluate its situational applicability in competition.
**Conclusion**: This study advanced the biomechanical understanding of Sutemi, Makikomi, and Tai Atari using two original theoretical models, highlighting the importance of Tori's body mass motion. The findings obtained can influence training and competition strategies, enhance technique effectiveness, and reduce injury risks.

## 1. Introduction

Sutemi-Waza (捨て身技) refers to techniques where Tori deliberately compromises equilibrium as a tactical application, a principle first outlined by Kano Jigoro Sensei (Kano 1986). These techniques involve high commitment and risk, often resulting in spectacular throws. Understanding neurological principles, such as motor redundancy, is crucial for appreciating the efficiency of these complex movements and their applications in Sutemi-Waza.

Motor redundancy (Schmidt 2011, Sinh et al. 2025), the ability of the nervous system to generate numerous solutions for a given task, is crucial for dual-situation sports such as judo.

This principle shows the importance of efficient technique execution by allowing the nervous system to coordinate redundant elements to optimise energy consumption. (Paillard 2010, Sacripanti 2016)

**Purpose and Significance**

This study examines the biomechanics governing Sutemi and Makikomi techniques recognised by the Kodokan (Kano 1986), presenting two original theoretical mathematical models. Understanding these techniques' biomechanics is vital to enhance effectiveness and ensure athlete safety. Despite their importance, a comprehensive biomechanical analysis, particularly related to Tori body mass and motion dynamics, remains understudied.

No scientific biomechanical publication deals organically with all Sutemi and Makikomi techniques as a whole, only sporadic studies on individual techniques' biomechanics (Lacharwar and alt. 2025, De Angelis 2020, Iura 1991) or biomechanical analysis of technique groups including some Sutemi (Imamura et alt. 2007, Hamaguchi 2024).

To address this gap, this study proposes two theoretical biomechanical models to fill the literature gap and provide coaches with insights into improving training strategies.

Furthermore, this study includes Tai Atari, a technical tool using body mass impact to throw opponents. Changes in refereeing regulations have enhanced the use of Tai Atari, a method rooted in ancient Samurai in Japan (Inogai 2018).

By connecting theoretical biomechanics and practical applications, this study explored the situational use and advantages of each technique group, offering insights for coaches, with data on technique utilisation in high-level male competition.

The results will deepen the biomechanical understanding of Sutemi, Makikomi, and Tai Atari, emphasising Tori's body mass dynamical motion.

Before examining methodology and results, it is essential to understand how biomechanics supports the effectiveness of judo techniques and guides our analysis.

For example, in Ma Sutemi Waza (back sacrifice techniques), biomechanical analysis helps understand how body weight distribution affects throws. It involves creating a perfect arc that uses the opponent's momentum during Tori's fall. Biomechanics highlights the importance of kinetic chain movements that generate and transfer energy through the body.

In Ma Sutemi Waza, initiating movement from the abdomen and coordinating through limbs ensures optimal efficiency and minimises exertion.

By understanding the qualitative mechanics, a coach equips an athlete to achieve the most harmonious outcome. Biomechanical knowledge helps athletes identify potential issues before injuries occur, allowing them to adjust movements and reduce injury risk (Penichet-Tomas A. 2025).

Sutemi Waza involves engaging with an opponent on the ground; thus, mastering descent control safeguards both athletes. It emphasises purposeful falling rather than collapse. These considerations underscore the significance of biomechanical analysis in judo projections and justify our methodological approach.

## 2. Methodology

This study employed biomechanical methodology to examine Sutemi and Makikomi techniques in judo. The investigation focuses on Tori's actions, exploring how Tai Atari influences technique execution and serves as an auxiliary tool in others, such as Ashi Waza throws (Sacripanti 2019).

 A central aspect was motor redundancy. In dual-situation sports like judo, motor redundancy enables the nervous system to generate multiple solutions for a task. This adaptability optimizes energy consumption of techniques by allowing efficient coordination of redundant elements. This principle is crucial for complex movements in Sutemi, Makikomi, and Tai Atari techniques, where precise timing and control are essential (Shing and alt. 2025). The approach began with analysing Ukemi Waza directions for all Sutemi and Makikomi classified by the Kodokan, for both Tori and Uke, to ensure correct movement and athlete safety. Subsequently, videos of athletes' technical actions in competition were analysed, examining body mass use during execution to identify biomechanical principles. The study was conducted through video observations, supplemented by technical discussions between the authors: Dr. Galea and Dr. Lascau, 8th Dan, and Professor Sacripanti, 7th Dan. This study allow us to develop original theoretical models representing Tori's body mass dynamics during Sutemi and Makikomi techniques.

And to formulate a general equation for the Tai Atari tool, highlighting optimal applicative methods for coaches' training and tactical implementation.

The following sections analyse these techniques, beginning with an overview, followed by the three-phase movement analysis with Kano Sensei's method, study of fall directions for safety, and development of three original biomechanical models describing Tori's body movement during execution.

## 3. Biomechanical Analysis of judo Techniques

### 3.1 Overview of Tori's Body Motion.

Building upon our understanding of judo's physical dynamics, this study examines Tori's body movements during Sutemi Waza execution, a technique requiring strategic balance sacrifice to throw Uke. These techniques are categorized by Kano Sensei and Kodokan Classification into Ma Sutemi waza, Yoko Sutemi waza, and Makikomi waza (Kano 1986), following biomechanical classification. All techniques belong to the lever group, with Sutemi classified under maximum arm and Makikomi under minimum arm subgroups (Sacripanti, 1987). Kano Sensei and his assistants identified three phases in throwing dynamics through preliminary biomechanical analysis. He introduced a foundational triad to judo pedagogy: Kuzushi (unbalancing the opponent), Tsukuri (positioning for the throw), and Kake (executing the throw).

 This triad serves as the fundamental building block of Nage Waza and Judo instruction. (Sacripanti 2014) No real temporal division exists between these phases; explicative artificiality allows teaching of a single, continuous movement.

The issue of precedence between the first two phases was resolved through Kodokan's early scientific research (Matsumoto 1978), which showed these phases were interconnected via electromyographic assessments. Thus, Kuzushi and Tsukuri phases are continuous movements that begin simultaneously. This division remains an instrumental system that facilitates teaching complex movements in simple steps.

The biomechanical analysis of Tori's Sutemi techniques involves three phases: 1. Kuzushi: disrupting Uke's balance to create an unbalanced position. 2. Tsukuri: positioning Tori's body by lowering his centre of gravity. 3. Kake: executing the throw. Tori employs rotational and translational movements using one leg as a fulcrum, applying force with his arms to project Uke's body (Sacripanti, 2014).

 The Kodokan recognizes 19 techniques (Daigo 2005): ten Sutemi, seven Makikomi, and two prohibited throws. The biomechanical analysis will examine these techniques to reduce the risk of injury, identifying the correct landing directions for both athletes, thus facilitating safer biomechanical models of body mass movement.

### 3.1.1 Sutemi-Waza (捨身技) - Sacrifice Techniques: Ukemi analysis

| Ukemi analysis Ma Sutemi Waza | Tori falls | Uke performs |
|---|---|---|
| 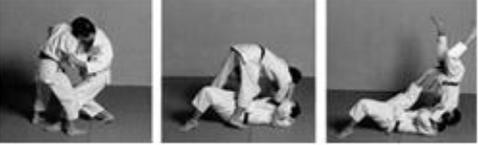 **Tomoe-nage** [1] | Ushiro-Ukemi | Mae-mawari-Ukemi |
| 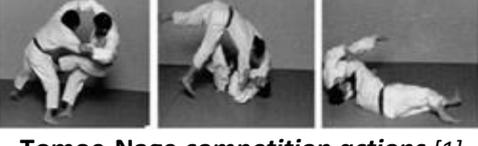 **Tomoe-Nage** *competition actions* [1] | Ushiro-Ukemi | Yoko-mawari-Ukemi (not Kodokan classified) |
| 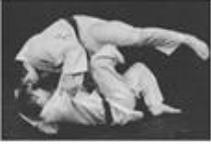 **Sumi-Gaeshi** [2] | Ushiro-Ukemi | Mae-mawari-Ukemi |
| 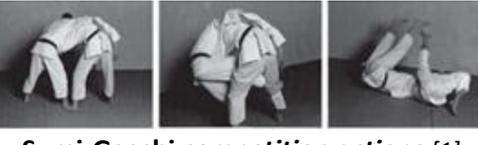 **Sumi-Gaeshi** *competition actions* [1] | Diagonal Ushiro-Ukemi | Yoko-mawari-Ukemi |
| 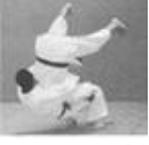 **Tawara-Gaeshi** [1] | Ushiro-Ukemi | Mae-mawari-Ukemi |
| 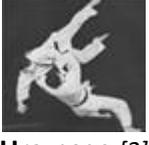 **Ura-nage** [2] | Ushiro-Ukemi | Mae-mawari- Ukemi |
| 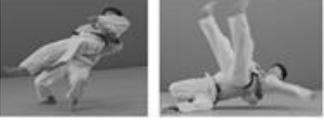 **Ura-nage** *competition actions* [1] | Diagonal Ushiro-Ukemi | Ushiro-Ukemi |
| **Footnote**: In the basic demonstration of Tomoe-nage, Sumi-gaeshi, Hikikomi-gaeshi, Tawara-gaeshi, and Ura-nage Tori falls the same Ushiro-ukemi, and Uke performs the same Mae-mawari-ukemi | | |
| **Note:** *The above technical illustrations were taken from five historical manuals, and the numbers in the subscripts indicate their provenances.* **[1]** *(Daigo2005)* **[2]** *(Kazuzo Kudo 1967)* | | |

| Ukemi analysis Yoko Sutemi Waza | Tori falls | Uke performs |
|---|---|---|
| 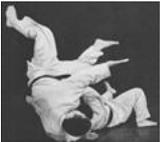 **Yoko-guruma** [2] | diagonally to Ushiro-Ukemi | diagonal mae-mawari-Ukemi. |
| 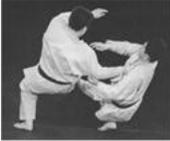 **Yoko-gake** [2] | Yoko-Ukemi | Yoko-Ukemi |
| 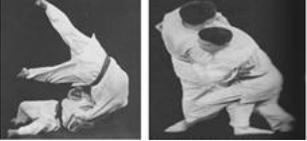 **Tani-otoshi** [2] | diagonally to the front of Yoko-kemi | Ushiro-Ukemi |
| 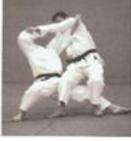 **Tani-otoshi variant, also known as *Waki-otoshi*** [1] | diagonally to the front of Yoko-ukemi | Ushiro-ukemi |
| 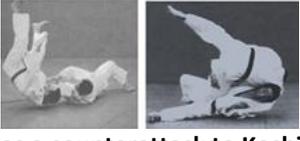 **Tani-otoshi, as a counterattack to Koshi-Guruma** [1] | Yoko-ukemi | Ushiro-ukemi |
| 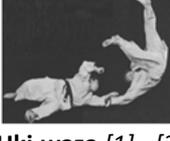 **Uki-waza** [1] - [2] | diagonally to Ushiro-ukemi | diagonal mae-mawari-ukemi (often falling on his side) |
| 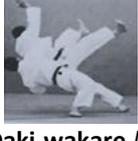 **Daki-wakare** [1] | diagonally to Ushiro-ukemi | diagonal Mae-Mewari-ukemi |
| 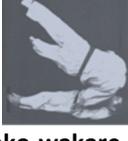 **Yoko-wakare** [1] | diagonal Ushiro-Ukemi | Mae-mawari-Ukemi |

***Note:*** *The above technical illustrations were taken from five historical manuals, and the numbers in the subscripts indicate their provenances.* **[1]** *(Daigo2005)* **[2]** *(Kazuzo Kudo 1967)*

| Ukemi analysis Makikomi-Waza | Tori falls | Uke performs |
|---|---|---|
| 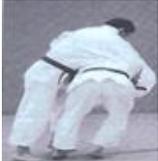 **Soto-makikomi** [1] | diagonal to Mae-Mawari-ukemi* | Mae-Mawari-ukemi |
| 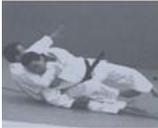 **Ko-uchi-makikomi** [1] | diagonal to Mae-Mawari-ukemi* | Ushiro-ukemi |
| 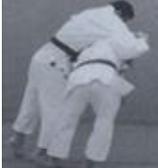 **Uchi-makikomi** [1] | diagonal to Mae-Mawari-ukemi* | Mae-Mawari-ukemi |
| 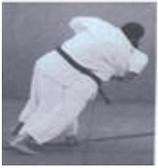 **O-soto-makikomi** [1] | diagonal to Mae-Mawari-ukemi* | Ushiro-ukemi |
| 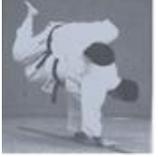 **Uchi-mata-makikomi** [1] | diagonal to Mae-Mawari-ukemi* | Mae-Mawari-ukemi |
| 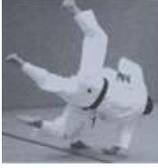 **Harai-makikomi** [1] | diagonal to Mae-Mawari-ukemi* | Mae-Mawari-ukemi |
| 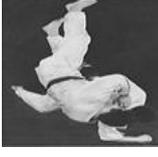 **Hane-makikomi** [2] | diagonal to Mae-Mawari-ukemi* | Mae-Mawari-ukemi |
| *Footnote makikomi: In Uchi-mata-makikomi, Harai-makikomi, and Hane-makikomi, Tori is the same, and Uke performs the same Ukemi* | | |
| 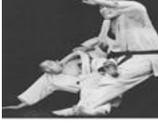 **Kani-basami** [2] | diagonally to Yoko-ukemi | Ushiro-ukemi |
| 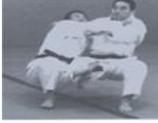 **Kawazu-gake** [1] | Ushiro-ukemi | Ushiro-ukemi |
| *Footnote IJF:* According current IJF refereeing rules Kani-basami and Kawazu-gake are forbidden in competition | | |
| *Note: The above technical illustrations were taken from five historical manuals, and the numbers in the subscripts indicate their provenances. **[1]** (Daigo2005) **[2]** (Kazuzo Kudo 1967)* *(often falling on his side due to the rotation stopped for safety by his arm on the tatami) | | |

**Not Kodokan Classified as Sutemi**

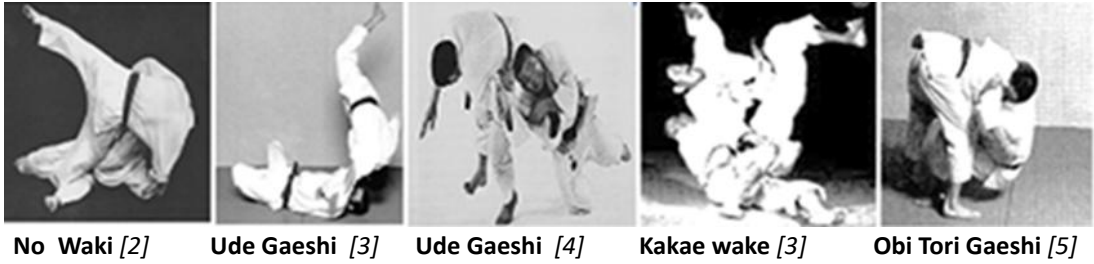

| No Waki [2] | Ude Gaeshi [3] | Ude Gaeshi [4] | Kakae wake [3] | Obi Tori Gaeshi [5] |

*Note: The above technical illustrations were taken from five historical manuals, and the numbers in the subscripts indicate their provenances.* **[2]** *(Kazuzo Kudo 1967)* **[3]** *(Mifune1961)* **[4]** *(Sato and Okano1973)* **[5]** *(Kashiwazaki 1980)*

After analysing the fall directions of Tori and Uke for Kodokan-approved techniques, which relate to athlete safety, we examined Tori's body motion to understand the mechanics and apply them correctly and effectively, providing useful information for coaches and athletes.

### 3.2 Biomechanical models of body motion in these techniques

#### A) Sutemi

In order to evaluate the Sutemi and Makikomi techniques, we have developed theoretical biomechanical models that simulate Tori's body motion during the application of Sutemi. The Ma Sutemi technique involves backward free fall of a body with increasing mass, as depicted in Fig. 1A, whereas the Yoko Sutemi technique entails side fall with increasing mass (Uke's mass), as illustrated in Fig. 1B. For the biomechanical analysis, we modelled the human body as a rigid cylinder (Sacripanti 2021). Both types of falls can be analysed similarly because the cylinders fall horizontally onto a surface. The distinction between the two techniques is addressed by considering rotary motion with angular momentum variation, considering the position of Tori's feet in both sliding and non-sliding scenarios. Here, h represents the body height, and M = m1 + m2 denotes the sum of the athlete's mass. Figure 1 illustrates a Tori's body (cylinder) falling horizontally.

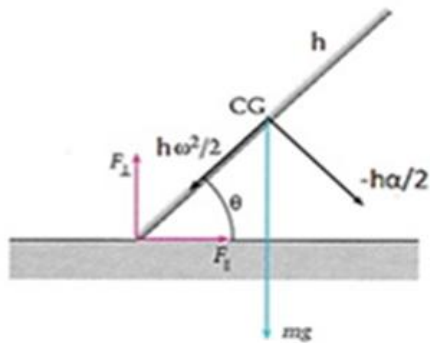 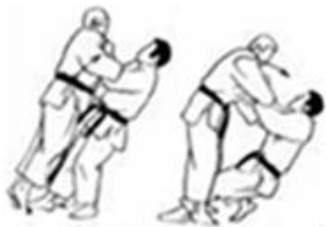 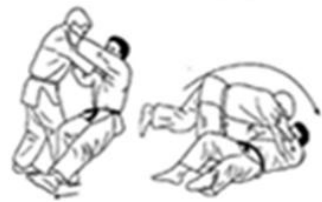

*Fig.1 Toris' body* (cylinder) *falling down*      *Fig.1 A   Ma Sutemi*      *Fig.1 B  Yoko Sutemi*

If ω and α are the angular velocity and acceleration of the body, respectively, we can relate the vertical and horizontal components of the Gravity Centre (CG) acceleration before the sliding onset to angles θ and α. The acceleration of the centre of gravity is influenced by the angular acceleration (-hα/2) and centripetal force ($\frac{h\omega^2}{2}$), as shown in Figure 1. If SL (sliding) and ST (stopping) represent directions parallel and perpendicular to the tatami, respectively, Figure 1 shows that the accelerations SL and ST are

$$\alpha_{SL} = -\frac{h\omega^2}{2}\cos\theta - \frac{h\alpha}{2}\sin\theta \qquad (1)$$

$$\alpha_{ST} = -\frac{h\omega^2}{2}\sin\theta + \frac{h\alpha}{2}\cos\theta \qquad (2)$$

Both are negative because no energy dissipation occurs before sliding begins.

The total energy of the two bodies falling together is

$$\tfrac{1}{6} M h^2 \omega^2 + \tfrac{1}{2} M g h \sin\theta = \tfrac{1}{2} M g h \qquad (3)$$

From this, we can evaluate both the square angular velocity and angular accelerations $\omega^2$ and $\alpha$:

$$\omega^2 = \tfrac{3g}{h}(1 - \sin\theta) \qquad (4)$$

$$\alpha = -\tfrac{3g}{2h}\cos\theta \qquad (5)$$

After performing calculations, we did not consider this study. It is possible to express the sliding and stopping forces, which will allow us, through their cases, to diversify the Tori motions associated with two classes: Ma and Yoko Sutemi (Hinrichsen 2021, Rod 2021, Kano 1986)

$$F_{ST} = \tfrac{Mg}{4}(1 - 3\sin\theta)^2 \qquad (6)$$

$$F_{SL} = \tfrac{3}{2}(Mg\cos\theta)(\sin\theta - 1) \qquad (7)$$

To understand how a judoka's body behaves during fall ( Laing 2010; van den Kroonenberg 1995; Pascoletti 2019), we can approximate this by summing up how each body part contributes to the overall movement by means of the Moment of Inertia. Knowing the exact mass and dimensions of each body segment is necessary for the accurate calculation of the total moment of inertia of the body ($I_{tot}$). In addition, the human body is flexible and rigid, which further complicates the calculations.

$$I_{tot} = I_{trunk} + I_{arm} + I_{leg}$$

where

$$I_{trunk} = \tfrac{1}{12} m_t \left(3 r_t^2 + h_t^2\right)$$

$$I_{arm} = \tfrac{1}{12} m_a \left(3 r_a^2 + h_a^2\right) \qquad (8)$$

$$I_{leg} = \tfrac{1}{12} m_l \left(3 r_l^2 + h_l^2\right)$$

where r is the half diameter of the trunk, arm, and leg, and h is the relative length of the body part. Using the two values of the moment of inertia of bodies, it is possible to diversify the motion of Ma Sutemi and Yoko Sutemi. In Ma Sutemi, techniques that are: Tomoe Nage, Sumi Gaeshi, Tawara Gaeshi, Hikikomi Gaeshi, Obi Tori Gaeshi, and Ura Nage
the (feet/foot) of Tori doesn't slip ($F_{SL}= 0$ ), but it's still; therefore: 1. $F_{ST} \neq 0$; and   $I_{tot} = I_{trunk} + I_{arms} + ½ I_{leg}$ (9)

       For Ura Nage:  $F_{ST} \neq 0$; and   $I_{tot} = I_{trunk} + I_{arms} + I_{leg}$ (10)

For all the Yoko Sutemi, the (feet/foot) of Tori is not still, then $F_{ST} = 0$

      2.$F_{SL} \neq 0$   and $I_{tot} = I_{trunk} + I_{arms} + I_{leg}$  (11)

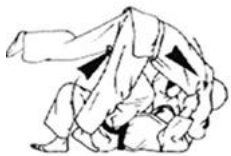 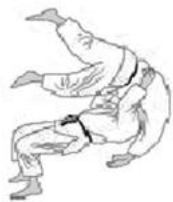 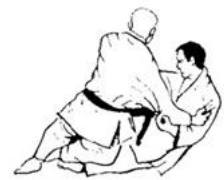

**Tomoe-Nage** equation (9)    **Ura-Nage** equation (10)    **Yoko-Gake** equation (11)

*Fig 2 Judo actions relative to the previous equations 9,10,11*

These conditions allowed us to evaluate the falling velocity, showing that the angular velocity and acceleration were proportional to the gravitational acceleration and body-Tatami angle and inversely proportional to the athlete's height. The stopping and sliding forces depend on the sum of the Tori and Uke masses and the body-Tatami angle but are independent of the athlete's height.

### B) Makikomi tool for direct attack or technical finishing

In judo, athletes' bodies are connected by their arms, which apply push-pull forces in all directions. The action of one body moving away from or toward another can be compared with the dynamics of a body subjected to elastic forces. According to Bertrand's theorem, only closed orbits—circular trajectories of motion—act under generalised elastic forces in judo movements, as in the dynamic clashes between Tori and Uke.

$$F = -kx^\alpha \quad (12)$$

Only for α = 1 and α = -2, as demonstrated by Goldstein (2000), Tori uses unique trajectories, called "General Action Invariants" (GAI) (Sacripanti 2010) to apply the lever or torque techniques defined earlier. (Sacripanti 1987) The next figure shows similar trajectories as Tori takes a contact point for a rapid internal rotation, applying the lever or torque to throw Uke. In terms of the elastic field, Tori's trajectories resemble the trajectory of the capture of one particle by another in the elastic field with internal forces.

$$F_{ab} = -F_{ba} = F \quad (13)$$

In this case, the two particles acted as athlete bodies during the inward rotation throw (Teodorescu 2009)

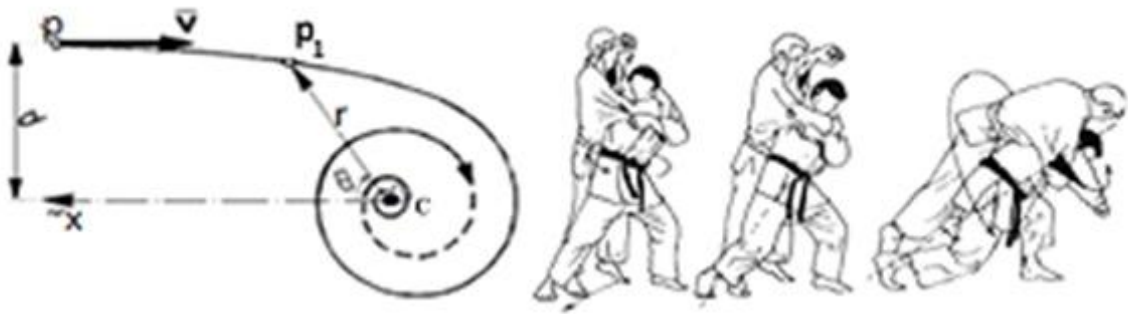

**Fig 3. The trajectory of a capture particle similar to Tori's throwing inward trajectory in judo Makikomi (Teodorescu 2009)**

*Almost-plastic collision of the extended bodies.*

The trajectory ends in a projection using a lever technique, starting with a collision (Tai Atari), which is considered almost plastic, as athletes are strictly connected, but bodies do not merge. When Tori and Uke collide, their bodies interact like a 'sticky' collision, moving together after impact. This 'almost-plastic collision' concept is crucial for understanding momentum transfer during throws. With equal mass between athletes, the equations were easy to obtain (Stronge, 2018). If they have different starting velocities, they remain connected until the fall, which is considered a free fall. With gravity force negligible at start but increasing after collision until Uke's landing, it is possible to write for early collision instants, as it is a rotational impact (Di Benedetto 2011)

$$mv_1 + mv_2 = 2mv \quad (14)$$

$$\text{the conservation of ang. momen. give us } I_1\omega_1 = (I_1 + I_2)\omega_f \quad 15)$$

$$or \frac{\omega_f}{\omega_1} = \frac{I_1}{(I_1+I_2)} \quad (16)$$

*the impact is totally inelastic, and the loss of kinetic Energy give us:*

$$\Delta K = \frac{1}{2}I_1\omega_1^2 - \frac{1}{2}(I_1 + I_2)\omega_f^2 = \frac{1}{2}\frac{I_1 I_2}{I_1 + I_2}\omega_1^2 \quad (17)$$

**Full rotation with free fall**

The complex rotational application of judo Makikomi connects to direct attack using lever techniques, with inward and outward rotation, or as finishing tactics after throws like Uchi Mata, Hane Goshi, Harai Goshi, Ko Uchi Gari. The physical model was a spinning top with variable rotational inertia. For Tori as an observer, the velocity transformation formula from the inertial frame to the rotating frame shows how speed is evaluated during rotating throws, as previously described (Sacripanti, 2014).

$$v = V + \left(\frac{dr'}{dt}\right)_O = V + (v' + \omega \wedge r') \quad (18)$$

It is important to evaluate the general equation of motion for this variable mass spinning up, remembering the classical Newton approach to rotational dynamics (Crabtree 1909) as follows:

The torque on the first athlete $\tau$ will be

$$\tau = r\,F \quad (19)$$

where $r$ is the radius between the centre of mass and the point at which the force $F$ is applied.

In terms of the rotational dynamics, this equation can also be written as

$$\tau = I\frac{d\omega}{dt} = mr^2\frac{d\omega}{dt} \quad (20)$$

To evaluate the general equation of motion for this variable mass spinning up, we recall the classical Newton's approach to rotational dynamics (Crabtree 1909):

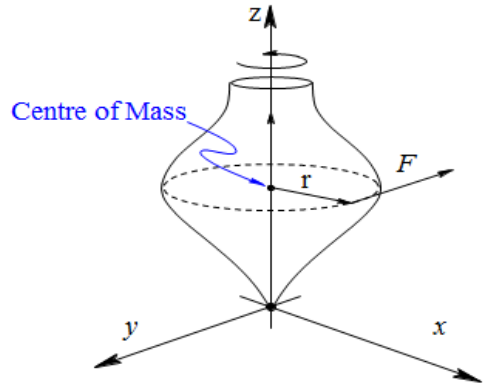

*Fig. 4 Spinning Top*

The Euler representation of a rotating rigid body (Tori) produces a nonlinear system of equations that are not always solvable analytically. The equations of motion for the Euler angle, as determined by Garanin (Garanin 2008), are as follows:

$$\begin{aligned}
\dot{\theta} &= \left(\frac{1}{I_1} - \frac{1}{I_2}\right)L\sin\theta\sin\psi\cos\psi \\
\dot{\varphi} &= \left(\frac{\sin^2\psi}{I_1} + \frac{\cos^2\psi}{I_1}\right)L \\
\dot{\psi} &= \left(\frac{1}{I_3} - \frac{\sin^2\psi}{I_1} - \frac{\cos^2\psi}{I_1}\right)L\cos\theta
\end{aligned} \quad (21)$$

Note that equations for θ˙ (nutation) and ψ˙ (spin) form an autonomous system that can be solved first; after that, the equation for precession can be integrated using the found spin to obtain the solution. With the Makikomi movement, the athlete tilts (Rose 2011), and the rotational axis through angle ϕ, as shown in Figure 5, becomes more complicated than that of a spinning top. The case of a tilted athlete applying Makikomi is shown in Figure 6, where gravity pulls down the centre of mass, pulling a non-spinning athlete downward, and increasing the tilt angle ϕ causes both athletes to fall. In a spinning top, the torque and change in the angular momentum vector are perpendicular to axis uˆ, leading the top to move "sideways" in a circle around the z-axis, called the precession. However, this phenomenon is nullified because of the increased mass of the system that connects the two athletes' falls after the almost-plastic collision.

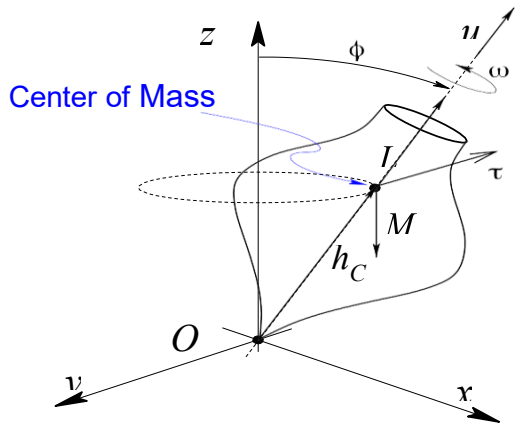
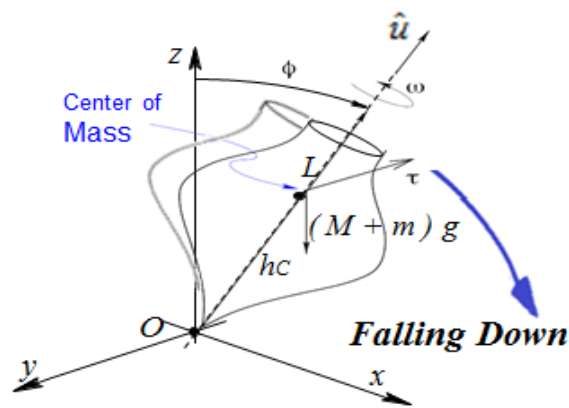

*Fig. 5*
*Tilting spinning top*

*Fig. 6*
*Variable Inertia Falling Spinning Top.* (Makikomi mechanics)

The three-dimensional equation becomes

$$\tau = r \wedge F \quad (22)$$

$$\tau = \frac{d(I\omega)}{dt} = \frac{d[(M+m)\omega]}{dt} \quad (23)$$

After contact, the mass increases, as in variable mass systems (Irschik 2014), the velocity decreases, and the external gravity force overcomes the precession motion, causing the body to drop. The Makikomi movement serves as a direct attack and tactical tool. This model showed that the required torque was proportional to the athlete's mass and Tori's angular velocity.

### C) Evolution Trends in Throwing
### the use of (Tai-Atari) [体(Tai): "body" 當 (Atari): "hit" or "strike"]
### An active method of using body mass by impact-to-throw.

In Sutemi Waza, Tori's body mass with hips, legs, and arms generates a throwing force without colliding with Uke, except at the fulcrum point. In Tai Atari, Tori used body mass to push opponents to the ground, which is common in modern competitions. These corollaries determine the directional force used in motion throws (Ikai, 1958; Sacripanti, 2015). Advanced Dynamic Imbalance: 1. Breaking body symmetry (Sacripanti 2014) shifts the COM, increasing the stability of Uke by 2. Timing fits the body's unbalanced position during collision 3. The forces remained effective in the 360° horizontal plane. These terms unify the biomechanical forces for the imbalance of uke during throws. Tai-Atari relies on body mass to overpower opponents. In modern judo: Positioning: Tori faces opponents. Impact: Tori uses mass to disrupt balance. Throw: Tori completes the technique. Macroscopic collisions involve energy dissipation. With adhesion, collisions can be perfectly inelastic with maximal energy loss (Preclik 2018). Athletes' direct impact is almost-plastic collision. When athletes grab during a collision, adhesion ensures same-speed movement. In inelastic collisions: (1) Total energy is conserved in inelastic collisions. (2) Kinetic energy loss (3) Linear momentum conservation (Goldstein 1960). For two-athlete direct-body collision, linear momentum expression remains same. However, kinetic energy expression is modified to

$$\tfrac{1}{2} m_1 u_1^2 + \tfrac{1}{2} m_2 u_2^2 = \tfrac{1}{2} m_1 v_1^2 + \tfrac{1}{2} m_2 v_2^2 + \delta \quad (24)$$

where δ is the positive kinetic energy dissipated in the collision between athletes. When judoka applies Tai Atari, second body is almost still before collision, and conservation of momentum and energy are modified as

$$m_1 u_1 = m_1 v_1 + m_2 v_2 \quad (25)$$

$$\tfrac{1}{2} m_1 u_1^2 = \tfrac{1}{2} m_1 v_1^2 + \tfrac{1}{2} m_2 v_2^2 + \delta \quad (26)$$

After calculation, from the conservation of momentum, because athletes' bodies stick together, the unknown velocity after collision is

$$v_2 = \left(\frac{m_1}{m_1+m_2}\right) u_1 \quad (27)$$

if $u_1$ is the "known" Tori attack velocity.

The maximum energy dissipated during the collision attack was.

$$\delta_{max} = \tfrac{1}{2} \left[\frac{(m_1 m_2)}{m_1+m_2}\right] u_1^2 \quad (28)$$

From these results, we understand that the attack velocity is the most critical parameter; the greater the attack velocity, the greater the energy dissipation during collision. If athletes have the same weight or mass M, the maximum energy dissipated will be

$$\delta_{max} = \tfrac{1}{4} M u_1^2 \quad (29)$$

Then, the final velocity for falling down the athletes' bodies, in this special case, will be

$$v_2 = \tfrac{1}{2} u_1 \quad (30)$$

With the inactive contrasting forces of the Uke, the applied power is greater, increasing both the maximum dissipation and the falling velocity. Our model shows that the final falling velocity and energy dissipation depend on the starting velocity of tori and are inversely proportional to the body masses with different mass numerators. With equal mass, the final velocity depends only on the initial attack velocity and is independent of body mass. The following figures (Sacripanti 2022) exemplify the Tai Atari attack with "Ashi Waza" techniques in competition, specifically: Ko Uchi Gake and O Uchi Gake

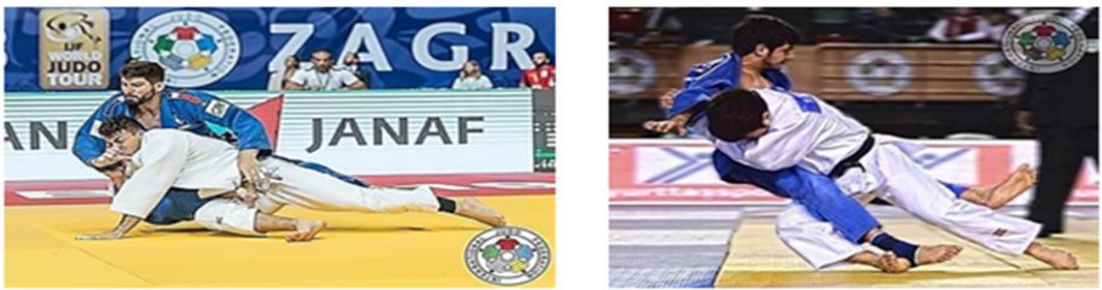

**Figg. 7-8 Right Ko Uchi Gake with Tai Atari (Sacripanti 2022 pag. 80)**

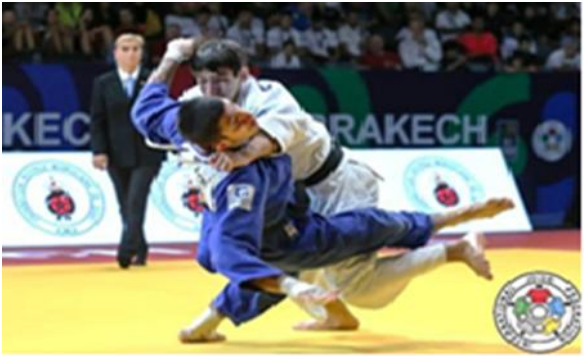 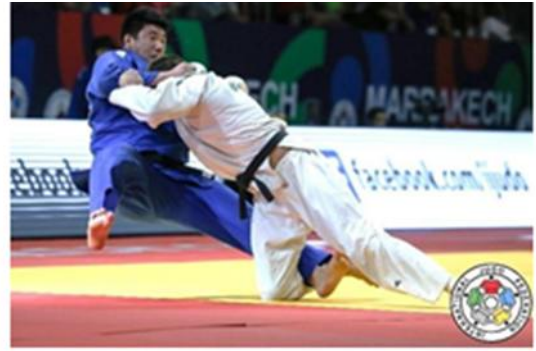

**Figg.9-10 Left and right O Uchi Gake with Tai Atari (Sacripanti 2022 pag. 80)**

After establishing the biomechanical underpinnings of the Sutemi, Makikomi, and Tai Atari techniques, we discuss their practical applications. This section translates biomechanical analysis into insights for coaches and athletes, focusing on strategies and their competitive advantages.

# 4. Results

**4.1 Effective Use of Body Mass in Sutemi and Makikomi Techniques**
Sutemi and Makikomi rely on the body mass of Tori.
**Sutemi Techniques**: Falling velocity, angular velocity, and acceleration are proportional to gravity acceleration and body angle with Tatami and inversely proportional to the athlete's body height.
The stopping and sliding forces depend on the combined masses of Tori and Uke and on Tori's body angle with the tatami.
**Makikomi Techniques**: Torque is proportional to the athletes' masses and to the angular velocity applied by Tori to Uke, underlining body mass application.
**4.1.1 General Equation for Tai Atari**
A general equation was defined for Tai Atari to assess its situational applicability.
Tai Atari leverages Tori's body mass during impact to facilitate the opponent's push to the ground.
This study also clarifies its competitive advantages and disadvantages.
**4.1.2 Energy Dissipation and Final Falling Velocity**
The final falling velocity and energy dissipation depend on Tori's starting velocity and are inversely proportional to the combined body mass. For athletes of equal mass, the final velocity depends only on the starting attack velocity.
**4.1.3 Statistical findings in high-level competition**
data were linked to the IJF weight categories (Sacripanti, 2021). Light categories (60-65 kg) preferred Sutemi techniques (11-16% effective rate), while heavyweight categories (100 to +100 kg ) favoured only Makikomi (5% intensity).

**4.2 The effectiveness of the Sutemi competition is based on simple informative summary.**
The authors examined Sutemi and Makikomi effectiveness (Ippon) by linking all Grand Slam events worldwide from October 2020 to May 2021. There is no significant variation in Sutemi's percentage use (Sterckowicz 2013; Sacripanti 2021; Sacripanti and Lascau in Preparation).
The research material included IJF video recordings of six Grand Slam events from October 2020 to May 2021, plus the Doha Masters held 11–13 January 2021.
The complete statistics, analysed with IBM SPSS Statistics software, will be presented in a forthcoming article (Sacripanti and Lascau in Preparation).
Here, we present a simple summary of the techniques' percentage use. Data were selected from male medallists (gold, silver, and two bronze medallists) and their opponents in all weight categories. The sample was based on seven competitions, totalling 947 bouts, including 196 medallists and their opponents who performed 800 successful throws (Ippon) and 127 Ne Waza victories. A total of 137 bouts were analysed: extra lightweight (-60 kg), 140 light middleweight (-66 kg), 138 lightweight (-73 kg), 141 middleweight (-81 kg), 136 middleweight (-90 kg), 135 middle heavyweight (-100 kg), and 120 heavyweight (+100 kg). The matches were knockout, quarterfinals, semifinals, repechage, third place, and final. The following tables provide an overview of the matches in all seven IJF competitions, divided by weight.

| Tab 1 global most important results about all seven IJF competition (Sacripanti & Lascau in preparation) | | | | | | | |
|---|---|---|---|---|---|---|---|
| Weight category Kg | Fight N. | 3 shido (%) | Golden Score (%) | Total Throws N. | Lever Group % | Couple Group % | Shortest fights % |
| -60 | 137 | 12 (8.7 %) | 27 (19.7%) | 120 | 63% | 37% | 8.7 |
| -66 | 140 | 33 (24,0%) | 36 (25.7%) | 100 | 64% | 36% | 12.1 |
| -73 | 138 | 16 (11.6%) | 27 (19.5%) | 124 | 41% | 59% | 10.8 |
| -81 | 141 | 28 (19.8%) | 25 (17.7%) | 110 | 59% | 41% | 13.4 |
| -90 | 136 | 28 (20.6%) | 26 (19.1%) | 105 | 52% | 48% | 19.1 |
| -100 | 135 | 15 (11.1%) | 23 (17,0%) | 141 | 51% | 49% | 15.5 |
| +100 | 120 | 15 (12.5%) | 16 (13.3%) | 100 | 52% | 48% | 21.6 |
| Tot | 947 | 147 (15.6%) | 180 (19%) | 800 | 434 (54%) | 366 (46%) | 136 (17%) |

For understanding throwing techniques and Biomechanical groups, the following table shows Japanese terminology utilisation of throws and their biomechanical groups

| Tab.2 Most Applied Effective Throwing Techniques in all seven Competitions ( Biomechanical group, Japanese name, and relative percentage) (Sacripanti & Lascau in preparation) | | | | |
|---|---|---|---|---|
| Biomec. Group | Japanese name | Total Number | Tot % | Group % |
| LEVER 54% 434 | Seoi Otoshi | 103 | 12.8% | 23.7% |
| | Sumi Otoshi | 79 | 9.8% | 18.2% |
| | Sumi Gaeshi | 30 | 3.7% | 6.9% |
| | Sode Tsuri | 29 | 3.6% | 6.6% |
| | Tai Otoshi | 23 | 2.8% | 5.3% |
| | other waza | 170 | 21,3% | 39,3% |
| | Total Group | 434 | 54% | 100% |
| COUPLE 46% 366 | Uchi Mata | 83 | 10.3% | 22.5% |
| | Ko Soto Gari | 70 | 8.7% | 19.1% |
| | O Uchi Gari | 67 | 8.3% | 18.3% |
| | Ko Uchi Gari | 61 | 7.6% | 16.6% |
| | O Soto Gari | 55 | 6.8% | 15,0% |
| | other waza | 30 | 4,3% | 8,5% |
| | Total Group | 366 | 46% | 100% |

Sutemi are preferred by lightweights (60-66 Kg), with 11–16% of the techniques applied. Heavy weights (over 90 kg) preferred Makikomi at approximately 5%. Middleweights (67-90 Kg) were applied to both Sutemi and Makikomi at 6–7%. The application frequency is low compared to Seoi Nage or Uchi Mata, which have 35–40% effectiveness. The high effectiveness of Sutemi in lightweights stems from its optimal leverage and momentum, as shown in the biomechanical analysis. Given these biomechanical benefits, further research could unlock competitive advantages. Understanding the biomechanical forces in sutemi allows coaches to train athletes to execute movements aligned with body mechanics, thereby reducing the risk of injury through proper alignment and force distribution. Future studies should quantify these advantages using empirical data analysis. The next section presents practical outcomes of biomechanical examinations of Sutemi, Makikomi, and Tai Atari in high-level judo competitions.

## 5. Discussion

Due to the lack of biomechanical models for Sutemi and Makikomi techniques, similar to those presented first in this study, and given the inadequate coverage of the topic in world scientific publication as stated in the introduction, the authors decided that, as developing this paragraph with normal content like: analysis of findings, comparison with previous studies, and future research directions was impossible, they will focus the discussion on strengths, weaknesses, findings analysis, and future research directions. Instead of comparative discussion, we present a more practical analysis for coaches and athletes, with implications for training and competition. This analysis was developed from our experiences as referees, competitors, coaches, and teachers.

For coaches, understanding these techniques' advantages and disadvantages helps because they are "extreme" techniques where executors risk everything for victory. These insights can help coaches recommend techniques based on match dynamics and opponent behaviour. See Table 3. The recommendations are presented as statements in a table format to facilitate better recall.

### 5.1 Finding analysis

In our biomechanical model of Sutemi related to optimal angular falling velocity depend on gravity acceleration and body angle, inversely relating to athlete height. These findings offer practical guidance for coaches training athletes. The study determined optimal torque for Makikomi, providing guidance on using body mass during execution to enhance performance. By understanding these biomechanics, athletes can better leverage body mass for powerful, efficient throws. The application of Tai Atari, where body mass collisions are focal, is highlighted for effectiveness in Makikomi and special throwing techniques (Ashi Waza) used in high-level competitions. The general equation for Tai Atari offers a fundamental resource for understanding when and how to use this technique strategically in competitions.

### 5.2 Strengths and weaknesses

The strengths of this work include the first comprehensive treatment of the subject from a biomechanical perspective and the original physical models. These, even with their inevitable simplifications, will not only guide new and more targeted research, but also provide essential quantitative data for coaches. Furthermore, despite their simplicity, the models themselves are useful for better understanding the inner mechanics of these techniques. It is interesting to underline that the strengths are, at the same time, also the weaknesses of this study; in fact, the models developed are, given the complexity of the competitive dynamics, extremely simplified, despite their originality, and can therefore provide only partially applicable indicative information to the competitive moment.

## 5.3 Practical Applications and Implications for Coaches and Athletes

| Tab 3 Coaching information: Coaching useful information about Sutemi and Makikomi | | |
|---|---|---|
| **Situational Use** | **Advantages** | **Disadvantages** |
| Ma Sutemi : Used when opponents push forward aggressively, utilizing their momentum.<br>Yoko Sutemi Waza: Effective when opponents resist, allowing force application from different angles.<br>Makikomi Waza: Safer techniques for referees to evaluate, usable as direct attacks or finishing tactics. (Sacripanti 2021) | - Surprise element against forward-pressing opponents<br>- Leverages opponent's momentum with less strength<br>- Versatile application in various positions<br>- Effective against defensive opponents<br>- Allows directional control<br>- Utilizes body weight effectively | - Risk of counterattack if executed incorrectly<br>- Requires precise timing and control<br>- Physical impact from falling<br>- Ground vulnerability if technique fails<br>- Complex execution challenges referee evaluation<br>- Requires high skill level<br>- Recovery time affects match pace |

## 5.4 Effectiveness of Tai-Atari

The effectiveness of Tai-Atari relies on the following biomechanical principles tab 4 and 5:

| Tab 4 Biomechanical principles in Tai Atari | | |
|---|---|---|
| **Centre of Gravity** | **Kinetic Chain** | **Leverage and Momentum** |
| Lowering centre of gravity before impact enhances stability and power. | Movement originates from feet through hips to torso, maximizing impact force. | Striking at the right angle uses opponent's momentum for effective throws. |

**General Movement**

| Tab. 5 Time steps in Tai Atari attack | |
|---|---|
| **Initial Engagement** | **Executing Impact** |
| **Grip**: Tori grips opponent's Gi for control.<br>**Feet Positioning**: Feet shoulder-width apart with bent knees for stability. | **Body Movement:** Tori drives forward using legs and hips. Impact can be rotational (as in Makikomi) or direct.<br>**Timing**: Impact should match opponent's movement for off-balance.<br>Transition to Throw:<br>**Follow-Up**: Tori transitions quickly to throwing.<br>**Control**: Maintain grip and body position throughout throw. |

### 5.4.1 Tactical Use in Competition     -

Tai Atari is a core component of competition, using body mass and its impact on tactical advantage. tab 6

| Tab.6 Miscellaneous information on Tai Atari tactical use | | | | |
|---|---|---|---|---|
| **Tactical Application**: | **Psychological Impact**: | **Footwork:** | **Timing:** | **Technique Refinement**: |
| Tai-Atari breaks an opponent's stance to create openings for complex techniques. (Gensoku No Genkei advanced Form) (Sacripanti 2022) | A well-executed Tai-Atari can intimidate opponents, disrupting their confidence.<br>Modern Adaptation: Tai-Atari focuses on efficiency in modern judo | Proper footwork is crucial for optimal positioning | Strike and throw timing is critical to catch opponents off guard. | Practice helps execute Tai-Atari smoothly, continuous and effective. |

**5.5 Future research**

Given these biomechanical and tactical benefits, further research could provide competitive advantages. Understanding biomechanical forces in Sutemi allows coaches to train athletes to execute movements aligned with body mechanics, reducing injury risk through proper alignment and force distribution. Future studies should quantify these advantages using empirical data analysis.

With the mechanical models presented here, future biomechanical research will benefit from a solid theoretical basis for exploring the topic. The transition to quantitative assessments will be more fruitful. The study of athlete safety will be facilitated by identifying the key mechanical parameters.

# 6. Conclusion

This study provides a comprehensive analysis of the biomechanics of Sutemi, Makikomi, and Tai Atari, emphasising the role of Tori's body mass and dynamics of physical lever application in executing these throws effectively.

By analysing specific techniques and their applications, this study extends our knowledge of judo biomechanics and offers insights for coaches and athletes. The goal is to enhance the effectiveness and safety of judo techniques, contributing to the sport's evolution and practitioners' success.

Our biomechanical analysis of Sutemi revealed that optimal angular falling velocity relates to gravity acceleration and body angle and is inversely related to athlete height. These insights guide athletes and coaches on technique effectiveness during competitions while minimising injury risk.

The study determined optimal torque for Makikomi techniques, providing guidance on utilizing body mass during execution to enhance performance. Understanding these biomechanical aspects helps athletes leverage body mass for powerful throws.

The application of Tai Atari, where body mass collisions are key, is effective in Makikomi and special throwing techniques (Ashi Waza) used in high-level competitions. The general equation for Tai Atari offers a resource for understanding when to use this technique competitively.

**Significance, Future Research and Weaknesses**

This research, with its mechanical models, contributes to judo biomechanics, providing insights into Sutemi, Makikomi, and Tai Atari techniques. These findings can influence training methods, enhance technique effectiveness, and reduce injury risk. Future research could explore biomechanical differences between practitioners and optimize techniques through targeted training. Applying these findings could help develop training programs to enhance athletic performance. Further investigations could examine the transferability of these biomechanical principles across martial arts to evaluate commonalities in technique execution. This will expand understanding of these principles' applications. The theoretical biomechanical models, given competitive dynamics complexity, are simplified; despite their originality, they provide general indicative information applicable to competition.